\documentclass[aps,prb,twocolumn,showpacs,preprintnumbers,amsmath,amssymb]{revtex4}

\usepackage{graphicx}
\usepackage{dcolumn}
\usepackage{bm}
\usepackage{color}
\begin{document}

\title{Pressure-Tuned Point-Contact Spectroscopy of URu$_2$Si$_2$ from Hidden Order to Antiferromagnetic States: Similarity of the Fermi Surface Gapping }

\author{Xin Lu, F. Ronning, P. H. Tobash, K. Gofryk, E. D. Bauer and J. D. Thompson }
\affiliation {Los Alamos National Laboratory, Los Alamos, New Mexico 87545, USA}


\begin{abstract}
We report soft point-contact spectroscopy studies of URu$_2$Si$_2$ both in the hidder order (HO) and the large-moment antiferromagnetic (LMAF) states accessed by pressure. In the HO state at ambient pressure, the spectroscopy shows two asymmetric peaks around the Fermi energy that emerge below the hidden order  temperature T$_{HO}$. In the LMAF state at higher pressures, the spectra are remarkably similar to those in the HO state, indicating a similar Fermi surface gapping in the HO and LMAF states and providing a new clue to unraveling the puzzling HO state.
\end{abstract}

                            
\maketitle

Many interesting and exotic ordered states emerge in  strongly correlated electronic systems. One example is the hidden order (HO) state of the heavy fermion material URu$_2$Si$_2$, which has remained mysterious ever since its discovery. Transport and thermodynamic measurements on URu$_2$Si$_2$ at ambient pressure have clearly shown a second-order phase transition at T$_{HO}$=17.5 K;\cite{TPalstra95PRLDiscovery, MBMaple86PRLpartiallygapped} however, no experiment has unambiguously identified the origin of the HO state despite considerable effort over the past 25 years. In the HO state, a tiny antiferromagnetic (AFM) moment of $\sim$0.03 $\mu_B$ has been detected, but now is generally believed to be caused by inhomogeneous stress or strain in samples and parasitic to HO.\cite{PGNiklowitz10PRLLarmor} Interestingly, superconductivity also emerges below T$_{sc}\sim$ 1.5 K in URu$_2$Si$_2$, coexisting with HO. Several theoretical models have been proposed to explain the nature of HO: spin or charge density wave, multipolar ordering,\cite{PSantini94PRLQuadrapolar, AKiss05PRBOctupolar, Kotliar09NatPhysHexadecapole} helicity order,\cite{Varma06PRLHelicityOrder} dynamical symmetry breaking \cite{TheoryElgazzar09NatMat, ThoeryOppeneer10PRB} and hybridization wave,\cite{Balatsky11PRLhybrizationwave} among others. However, no consensus has yet been reached.\cite{Mydosh-Oppeneer-arXiv}

URu$_2$Si$_2$ displays a rich phase diagram under pressure:\cite{PGNiklowitz10PRLLarmor, EHassinger08PRBPTphasediagram, PressureNPButch10PRB} The HO transition temperature T$_{HO}$ slightly increases with pressure while superconductivity is suppressed and finally disappears at a low critical pressure P$_x \sim$ 0.5 GPa. At low temperatures near  P$_x$ , there is a first order transition from the HO to a  large-moment antiferromagnetic (LMAF) state with a moment of $\sim$0.4 $\mu_B$ and wavevector $Q_{AF}$=(0, 0, 1). The HO and LMAF phase boundary $T_x(P)$ meets the T$_{HO}$ line at a bicritical point (T$_c\sim$ 19 K, P$_c \sim$ 1-1.36 GPa) and, above P$_c$, LMAF order emerges directly from the paramagnetic state below T$_N$. While the HO and LMAF states are different states, they share many remarkable similarities in their transport and thermodynamic properties, \cite{MMcElfresh87PRBrhovsPressure, JJeffries07PRLpressureorder, EHassinger08PRBPTphasediagram} indicating an intimate relationship between them. Furthermore, Shubnikov-de Haas measurements also show that the Fermi surface does not change dramatically between HO and LMAF states.\cite{dHvaHassinger10PRL}

Recent inelastic neutron scattering measurements find that a longitudinal spin fluctuation in the HO is frozen into static AFM moments in the LMAF state and they have the same commensurate wavevector Q$_{AF}$=(0, 0, 1). \cite{NeutronVillaume08PRB, NeutronFBourdarot10JPSJ} It is thus argued that this commensurate spin resonance is a signature of the hidden order state. On the other hand, an incommensurate spin gap with wavevector Q$_1=(1\pm0.4$, 0, 0) persists from the HO to LMAF state with an increase in gap energy above P$_x$.\cite{NeutronFBourdarot10JPSJ} Also from neutron scattering experiments, Wiebe \textit{et al.} \cite{Wiebe07NatPhyNeutron} argue that the gapping of the incommensurate spin excitations in the HO state can explain the entropy loss below T$_{HO}$ and thus plays an important role in the formation of HO. 

Though some properties of URu$_2$Si$_2$ remain controversial, it is well established that a partial Fermi surface (FS) gapping with an associated reduced carrier number occurs below T$_{HO}$ and it continues even in the LMAF state. \cite{MBMaple86PRLpartiallygapped, JJeffries07PRLpressureorder, NeutronFBourdarot10JPSJ, YSOh07PRLHallEffect} As pointed out,  gapping of the spin-excitations could be a natural consequence of a particle-hole condensation with gapping of the electronic spectrum below T$_{HO}$. \cite{Balatsky09PRBSpinResonance} Spectroscopic studies that probe the gap as a function of pressure are lacking but are desirable to understand the nature of the FS gapping and its evolution with pressure. Scanning tunneling spectroscopy (STS) measurements have revealed a Fano line shape in the density of states (DOS) below the coherence temperature T$^*$ but well above T$_{HO}$. A mean-field gap-like feature develops on cooling below T$_{HO}$, indicating HO as a modification of the hybridization process at  T$^*$. \cite{STMJCDavis10, STMYazdani} Angle-resolved photoemission spectroscopic (ARPES) studies claim a heavy band appears near the Fermi energy (E$_F$) right at the HO transition.\cite{ARPESSantanderSyro09, ARPESYoshida10PRB} These powerful tools, however, can not be applied at pressures where LMAF develops. Soft point-contact spectroscopy (SPCS), easily adapted to a high pressure environment,\cite{VASidorov} serves as a unique tool to explore the electronic structure around E$_F$ in both the HO and LMAF states. 

In this article, we report the first SPCS measurements on URu$_2$Si$_2$ under nearly hydrostatic pressures that tune URu$_2$Si$_2$ from the HO to LMAF state. Spectroscopies of the two states display strikingly similar features: a symmetric peak centered around E$_F$ when $T_{HO} (T_N) < T <T^*$, resulting from the hybridization between conduction and 5f electrons. Two asymmetric peaks emerge below the transition temperature T$_{HO}$ (T$_N$), indicating the persistence of FS gapping from HO to LMAF states. These results further constrain possible theoretical models for the origin of the HO. 

URu$_2$Si$_2$ single crystals were grown by the Czrochralski method and electro-refined in vacuum at 1225 $^o C$ for 3 weeks. The residual resistance ratio (RRR=$\rho_{300K}/\rho_{0}$) is $\sim$375, indicating high crystal quality. Instead of the conventional point-contact method of engaging a sharp metallic tip on the sample, soft point-contacts were made by dipping the end of a 25 \textit{$\mu$m}-diameter platinum wire into Ag epoxy and attaching it to the surface of the crystal. This method proves to be effective to study the order parameters (OP) of various superconductors (Ref. \onlinecite{Gonnelli10SPCSReview} and references within), with its advantages of reliable stability over a large temperature range. Here we extend this soft point-contact technique to a pressure-dependent study on the behavior of URu$_2$Si$_2$ in its HO and LMAF states. The crystal was mounted in a BeCu/NiCrAl hybrid clamp-type pressure cell with silicone fluid as the pressure transimitting medium, which provides a very nearly hydrostatic environment. The pressure at low temperatures was determined from the resistively measured change in the superconducting transition temperature of Pb. The differential conductance G=dI/dV as a function of bias voltage V was recorded by a standard lock-in technique, with the sample biased positively for all the measurements. The contact resistance R around T$_{HO}$ is usually around 10 $\Omega$ and we can estimate the contact radius \textit{d} according to the Wexler formula $R=(4\rho l/3\pi d^2)+\Gamma (l/d) \rho/2d$,\cite{Wexlerformula} where $\rho$ is the resistivity and \textit{l} is the electron mean free path of URu$_2$Si$_2$ at low temperatures. If we take $\rho \sim$ 40 $\mu \Omega$ cm and \textit{l}$\sim$ 100 \AA,\cite{PCSURu2Si2Escudero94PRB} the estimated contact radius \textit{d} is 250 \AA. The contact, however, is made by hundreds of silver particles in parallel, each with average contact radius \textit{d$_0$}. Conservatively, 100$\times d_0 \sim 250$ \AA \ implied $d_0 \ll l$. This estimate, along with a total contact resistance at higher voltage bias that does not shift with increasing temperature up to 60 K, ensures the ballistic nature of the contacts.

\begin{figure}[http]
\includegraphics[angle=0,width=0.46\textwidth]{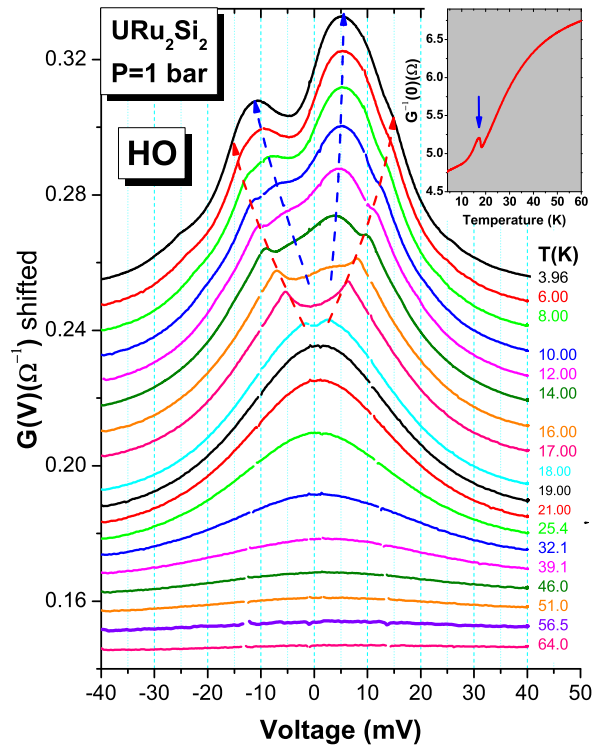}
\vspace{-13pt}
\caption{\label{fig:PCSHO} (color online). Differential conductance versus voltage curves, G(V), of  URu$_2$Si$_2$ at atmospheric pressure.  A soft point-contact is made on the c-axis. Curves at T $\leq$ 56.5 K have been shifted vertially for clarity and the corresponding temperatures are indicated next to each curve. The dashed lines are guides to the eye to show the evolutions of the side feature (red) and asymmertic peaks (blue). Inset: the temperature-dependent zero-bias contact resistance G$^{-1}_0$(T). The arrow indicates the local resistance maximum related to the side feature of G(V).}
\end{figure}

Figure \ref{fig:PCSHO} shows one set of representative spectra G(V) for URu$_2$Si$_2$ at ambient pressure and various temperatures from 64 K in the paramagnetic state to 3.96 K deep inside HO. At 3.96 K, two asymmetric peaks are clearly present. From multiple spectra taken with over ten contacts on different samples, the negative-biased peak always has a smaller conductance amplitude than the positive-biased one and the peaks are located at  $\Delta_0^- = -11\pm2$ meV and $\Delta_0^+=6\pm1$ meV, respectively. These observations are consistent with previous conventional mechanical PCS reports.\cite{PCSURu2Si2Escudero94PRB, PCSSamuely95PhysB, PCSRodrigo97PRB, PCSNaidyuk01LowTemp} Even though the conductance peaks observed here are asymmetric and offset with respect to E$_F$, they are very similar to results of point-contact Andreev-reflection studies of superconductors (Ref. \onlinecite{Gonnelli10SPCSReview} and references within), possibly indicating a mean-field like particle-hole condensation and macroscopic coherence build up in the HO state of URu$_2$Si$_2$. Based on this similairty, we simply assume the peak to peak distance is twice the charge gap $\widetilde{\Delta}_0$ of the HO state and estimate $\widetilde{\Delta}_0 =8.5 \pm 2$ meV, which is consistent with the gap estimated from resistivity,\cite{JJeffries07PRLpressureorder, NeutronFBourdarot10JPSJ} far-infrared reflectance \cite{DABonn88PRLInfrared} and specific heat measurements.\cite{MBMaple86PRLpartiallygapped}  

The conductance spectra at 64 K, well above T$_{HO}$, are flat; however, a nearly symmetric peak emerges around zero-bias voltage below the coherence temperature $T^*\sim 55$ K of URu$_2$Si$_2$, where the conduction electron bands and U's  5f-electrons begin to hybridize and the electronic structure of URu$_2$Si$_2$ near E$_F$ is modified. These spectra, though, are not typical of an asymmetric Fano-like shape that is found in STS measurements on URu$_2$Si$_2$ above T$_{HO}$.\cite{STMYazdani, STMJCDavis10} The absence of a Fano-like lineshape in G(V) plotted in Fig.~\ref{fig:PCSHO} may be a consequence of the much broader area and associated averaging of tunneling processes in these soft point-contact measurements.

\begin{figure}[http]
\includegraphics[angle=0,width=0.46\textwidth]{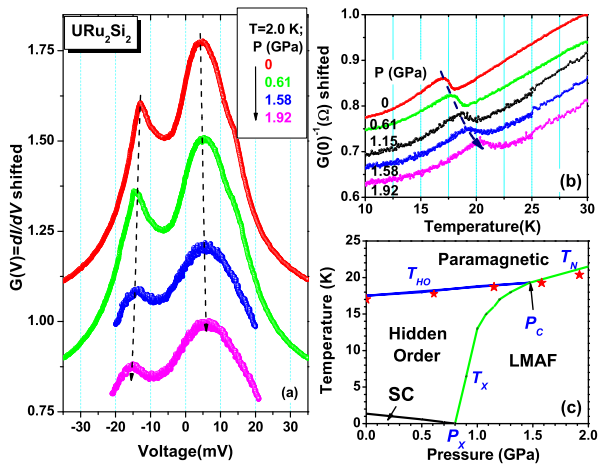}
\vspace{-13pt}
\caption{\label{fig:PCSPressure} (color online). (a)  Soft point-contact conductance spectra G(V) of URu$_2$Si$_2$ at T=2.0 K under different pressures. The dashed lines are guides to eyes, indicating a weak increase of the peak positions with pressures while the asymmetric peaks persist into the LMAF state. (b) Temperature-dependent zero-bias contact resistance curves G$^{-1}_0(T)$ at different pressures. The curves are shifted vertically for clairity and the dashed line is a guide to eye to indicate the evolution of local resistance maxima under pressure. (c) The pressure phase diagram of URu$_2$Si$_2$ taken from Ref.~\onlinecite{PressureNPButch10PRB}. The local zero-bias resistance maxima (stars) under pressure follow the pressure-dependent T$_{HO}$ (T$_N$) line.}
\end{figure}

The inset of Fig.~\ref{fig:PCSHO} shows the zero-bias resistance of the point-contact G$^{-1}(0)$ on URu$_2$Si$_2$ as a function of temperature, which has a jump around T$_{HO}$ similar to its conventional resistively measured behavior. With the good temperature stability of SPCS, we can track the temperature evolution of conductance curves G(V), especially at temperatures around $T_{HO}$. The jump in G$_0^{-1}$(T) corresponds to a side feature in G(V) present below 18.0 K and evolving to larger bias voltages with decreasing temperature. At 3.96 K, the side feature is dominated by the aymmetric peaks and can barely be detected. Because the resistivity jump is associated with the FS reorganization and thus a sudden removal of parts of the FS at the HO transition,\cite{EHassinger08PRBPTphasediagram,ThoeryOppeneer10PRB} the observed side feature in G(V) should be intrinsic to URu$_2$Si$_2$ and could be due to a change in the electronic DOS as the FS reconstructs. Previous conventional PCS measurements have reported that the onset temperature of PCS features differs from T$_{HO}$,\cite{PCSSamuely95PhysB, PCSRodrigo97PRB} possibly due to a local pressure induced at the point-contact area, which may affect properties locally.\cite{PCSRodrigo97PRB} On the other hand,  the appearance of a pseudogap due to incoherent fluctuations of the HO  has been proposed to explain observations of gap-like features several degrees above T$_{HO}$. \cite{JHaraldsen10Pseudogap, MKLiu11PRBHOPseudogap} Our SPCS measurements with different contacts on pristine samples do not show such a behavior. However, after pressure cycling our sample, the onset temperature of side features in G(V) rises to $\sim$25 K, far above T$_{HO}$ or T$_N$ for all new contacts. This suggests that stress or strain intrinsic to a sample or induced, e.g. by non-uniform pressure under a conventional point-contact tip, could play a role in producing the possible pesudogap phenomena in URu$_2$Si$_2$.

We have measured the point-contact spectra under a series of pressures tuning URu$_2$Si$_2$ from the HO to LMAF state up to 2 GPa and Fig. \ref{fig:PCSPressure} (a) displays the conductance curves G(V) at T=2.0 K for representative pressures. The point-contact resistance usually decreases somewhat under pressure and, consequently, the bias voltage spans a smaller range at higher pressures due to  limitations of the bias current. Nevertheless, the asymmetric conductance peak structure remains robust over the whole pressure range from HO to LMAF and the peak positions increase only slightly. Fig. \ref{fig:PCSPressure} (b) shows the evolution of the temperature-dependent zero-bias contact resistance G$^{-1}$(0) under different pressures. The peak structures persist from HO to LMAF states and the local maximum moves to higher temperatures, tracking T$_{HO}$ (T$_N$) in the pressure phase diagram shown in Fig. \ref{fig:PCSPressure}(c).

\begin{figure}[http] 
\includegraphics[angle=0,width=0.46\textwidth]{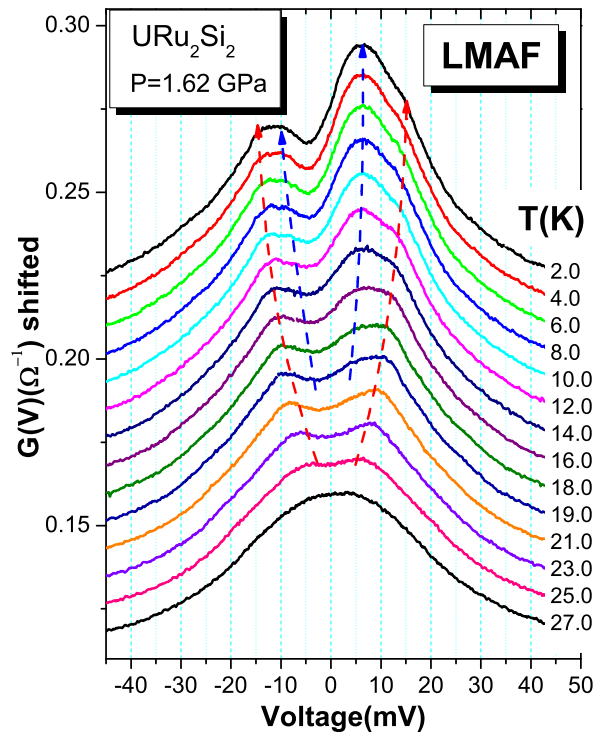}
\vspace{-13pt}
\caption{\label{fig:PCSLMAF} (color online). Differential conductance versus voltage curves G(V)  for URu$_2$Si$_2$ at P=1.62 GPa, with LMAF order as the ground state at this pressure. Curves at T$\leq$ 25.0 K have been shifted vertially for clarity and the corresponding temperatures are labeled on the right. The dashed lines are guides to the eye to show the evolution of the side feature and asymmetric peaks.}
\end{figure}

Although there are differences in reported T-P phase diagrams, especially in the exact pressure for the HO-LMAF phase boundary, \cite{PGNiklowitz10PRLLarmor} our point-contact measurements at P=1.62 GPa and T=2.0 K shown in Fig. \ref{fig:PCSLMAF} are well inside the LMAF state, independent of the details of the phase boundary.  An important conclusion from these measurements is that the asymmetric peaks, side feature in the conductance and temperature evolution of G(V) are still present and virtually unchanged relative to the HO state.  

The asymmetric peaks in PCS data at ambient pressure can be interpreted in either localized or itinerant pictures. For instance, Rodrigo \textit{et al.} argue \cite{PCSRodrigo97PRB} that, for point-contact measurements, the ``Kondo like'' resonance observed above T$_{HO}$ is split into two asymmetric peaks by the appearance of quadrupolar ordering below T$_{HO}$; however, no signatures of crystalline electric field splitting, expected in a localized picture,  have been observed. In an itinerant picture, asymmetric conductance peaks are anticipated in the tunneling density of states as a result of a hybridization gap, provided the scattering rate is sufficiently low. \cite{PColeman09PRLKondolattice, Wolfle10PRLKondolattice} Experimentally, however, the asymmetry manifests itself just below the second-order phase transition temperature T$_{HO}$ (T$_N$), indicating an unconventional form of hybridization if this scenario stands (one example is the hybridization wave, as in Ref. \onlinecite{Balatsky11PRLhybrizationwave}). On the other hand, neutron scattering \cite{Wiebe07NatPhyNeutron} finds a gap in inelastic spin excitations developing at the incommensurate wavevector Q$_1$  as URu$_2$Si$_2$  is cooled below T$_{HO}$ and this excitation accounts for much of the entropy change associated with HO.  In an itinerant picture, the strong coupling of spin and charge degrees of freedom should lead to a gap of similar magnitude in charge excitations, which is indeed consistent with the size of the asymmetric peak splitting we observe at low temperatures.

We now consider the implications for the robustness of the asymmetric peaks as a function of pressure. In an itinerant picture, the fact that SPCS data show that their energy scale does not change even deep into the LMAF state implies a similar charge and spin gap as at ambient pressure. Indeed, neutron scattering measurements under pressure continue to observe the gapping of inelastic spin excitations at Q$_1$, while concomitantly observing the magnetic order that forms at Q$_{AF}$.\cite{NeutronFBourdarot10JPSJ} The change in energy scale of these incommensurate excitations when LMAF develops at high pressures implies that excitations at Q$_1$ are coupled to the magnetic OP which is manifest at Q$_{AF}$. The fact that the Fermi surface is unchanged in the HO and LMAF states \cite{dHvaHassinger10PRL} equally implies that the excitations at Q$_1$ should be coupled to the OP at Q = (0, 0, 1) in the HO state at low pressures.\cite{MPezzoli11PRB} The microscopic origin for such a coupling is unknown. Though a theoretical model \cite{Kotliar09NatPhysHexadecapole} that treats the HO and LMAF states as a single complex OP $\Phi$ could account for robustness of asymmetric peaks in both HO and LMAF states as a trivial result of the rotation of a single OP as a function of pressure, this model relies on the presence of crystal fields, which have not been observed. Thus, this and other theories that involve crystal field effects, though attractive, appear to have their own shortcomings, which may be resolved in future experiments.

In conclusion, we have extended the soft point-contact spectroscopy technique under pressure to make the first charge-spectroscopy measurements of URu$_2$Si$_2$ in both HO and LMAF states. The asymmetric peaks observed in the HO state persists in the LMAF state and are characteristic of a charge gap around E$_F$ in the HO (LMAF) state. The persistence of the asymmetric charge gaps around E$_F$ in URu$_2$Si$_2$ points to a common origin for the charge gap in both HO and LMAF states, which constrains a theoretical interpretation of pressure-induced evolution of these states in URu$_2$Si$_2$.

We are grateful to A.V. Balatsky, W. K. Park, M. J. Graf, T. Park, V. A. Sidorov and J. X. Zhu for valuable discussions. Work at Los Alamos was performed under the aupices of the U. S. Department of Energy, Division of Materials Science and Engineering and suported in part by the Los Alamos LDRD program.

\end{document}